\documentclass[iop]{emulateapj}
\usepackage{apjfonts}
\usepackage{amsmath}
\usepackage{hyperref}
\usepackage{natbib}
\begin{document}
\shorttitle{SHAPES OF INTRACLUSTER GAS}
\slugcomment{{\em The Astrophysical Journal, in press}} 
\shortauthors{Lau et al.}

\title{Shapes of Gas, Gravitational Potential and Dark Matter in $\Lambda$CDM Clusters}

\author{
Erwin T. Lau\altaffilmark{1,2}, 
Daisuke Nagai\altaffilmark{3,4},
Andrey V. Kravtsov\altaffilmark{1,5},  
and Andrew R. Zentner\altaffilmark{6}
}

\altaffiltext{1}{Department of Astronomy \& Astrophysics, 5640 South Ellis Ave., The
  University of Chicago, Chicago, IL 60637 ({ethlau@oddjob.uchicago.edu})}
\altaffiltext{2}{Key Laboratory for Research in Galaxies and Cosmology, Shanghai Astronomical Observatory; The Partner Group of MPA; 80 Nandan Road, Shanghai 200030, China}  
\altaffiltext{3}{Department of Physics, Yale University, New Haven, CT 06520}
\altaffiltext{4}{Yale Center for Astronomy \& Astrophysics, Yale University, New Haven, CT 06520}
\altaffiltext{5}{Kavli Institute for Cosmological Physics and 
  Enrico Fermi Institute, 5640 South Ellis Ave., The University of
  Chicago, Chicago, IL 60637}
\altaffiltext{6}{Department of Physics and Astronomy, University of Pittsburgh, Pittsburgh, PA 15260}

\begin{abstract}
We present analysis of the three-dimensional shape of intracluster gas
in clusters formed in cosmological simulations of the $\Lambda$CDM
cosmology and compare it to the shape of dark matter (DM) distribution and
the shape of the overall isopotential surfaces.  We find that in
simulations with radiative cooling, star formation and stellar
feedback (CSF) intracluster gas outside the cluster core ($r\gtrsim
0.1r_{500}$) is more spherical compared to non-radiative (NR)
simulations, while in the core the gas in the CSF runs is more
triaxial and has a distinctly oblate shape. The latter reflects the
ongoing cooling of gas, which settles into a thick oblate ellipsoid as it
loses thermal energy. The shape of the gas in the inner regions of
clusters can therefore be a useful diagnostic of gas cooling. We find
that gas traces the shape of the underlying potential rather well outside the core, as expected in hydrostatic equilibrium. At
smaller radii, however, the gas and potential shapes differ
significantly. In the CSF runs, the difference reflects the fact that
gas is partly rotationally supported.  Interestingly, we find that in
non-radiative simulations the difference between gas and potential
shape at small radii is due to random gas motions, which make the gas
distribution more spherical than the equipotential surfaces. Finally,
we use mock {\sl Chandra} X-ray maps to show that the differences in
shapes observed in three-dimensional distribution of gas are
discernible in the ellipticity of X-ray isophotes. Contrasting the
ellipticities measured in simulated clusters against observations can
therefore constrain the amount of cooling in the intracluster medium and the presence of
random gas motions in cluster cores.
\end{abstract}



\section{Introduction}
\label{section:intro}

In the prevailing, hierarchical cold dark matter (CDM) paradigm of
cosmological structure formation, galaxy- and cluster-sized CDM halos
are formed via accretion and merging with smaller halos.  The CDM
paradigm predicts that DM halos are generally triaxial and
are elongated along the direction of their most recent major mergers.
The triaxiality of DM halos has been demonstrated in a number
of studies using numerical simulations
\citep{frenk_etal88,dubinski_carlberg91, warren_etal92,
  thomas_etal98,jing_suto02,suwa_etal03,hopkins_etal05,kasun_evrard05,
  allgood_etal06, bett_etal07, gottloeber_yepes07,paz_etal08} and arises due to 
anisotropic accretion and merging along filamentary structures. The
degree of triaxiality strongly correlates with the halo formation time
\citep[e.g.,][]{allgood_etal06,ho_etal06,wray_etal06}, which implies
that at a given epoch more massive halos are more triaxial. For the
same reason, triaxiality is sensitive to the linear structure growth
function and is higher in cosmological models in which halos form more
recently \citep[e.g.,][]{maccio_etal08}. 

Although shapes of DM halos have been studied extensively in 
dissipationless $N$-body cosmological simulations, DM shape is
difficult to probe observationally, though some handle on shape is provided by 
lensing studies
\citep[e.g.,][]{hoekstra_etal04,parker_etal07,rozo_etal07,evans_bridle09,hawken_bridle09}. 
Moreover, it is well known that including baryons in simulations modifies the
shapes of DM halos, especially in the case of significant gas
dissipation during galaxy formation \citep[][see
  \citeauthor{debattista_etal08} \citeyear{debattista_etal08} and
  \citeauthor{valluri_etal10} \citeyear{valluri_etal10} for discussion
  of the physical nature of this
  effect]{katz_gunn91,evrard_etal94,dubinski_94,kazantzidis_etal04,
  springel_etal04, hayashi_etal07,tissera_etal10}. 
It is therefore of paramount importance to examine predictions for halo shapes 
using cosmological simulations that include gas dynamics and dissipative
processes accompanying galaxy formation. Further, the shape of the
gas itself can be examined in such simulations and compared to the
shape of the underlying potential, sourced predominantly by DM.  
Gas is expected to follow isopotential surfaces in hydrostatic equilibrium,
so simulations may test whether cluster gas is in equilibrium
on average and whether it can be used as a reliable tracer 
of the shape of the underlying potential 
\citep{buote_tsai95,lee_suto03,flores_etal07,kawahara10}.

Probing the shape of the gravitational potential via gas, as was first
suggested by \citet{binney_strimpel78} and observationally tested by
\citet{fabricant_etal84} and \citet{buote_canizares96}, can open interesting avenues for using the
shapes of DM halos around observed galaxy clusters to both
test the CDM paradigm and constrain the amount of halo gas that
dissipated and was converted into stars during halo formation. This is
particularly relevant for galaxy clusters, where high-quality X-ray
imaging data now exists for large samples of clusters.
\citet{kawahara10} recently analyzed axis ratios of X-ray clusters
from the {\em XMM-Newton} catalogue and found relatively good
agreement with the CDM predictions of \citet{jing_suto02} based on
dissipationless simulations of a large cluster sample, confirming
findings of \citet{buote_tsai95} and \citet{flores_etal07} based on a
single simulated system.\footnote{A similar test of the shape
  determined using the observed projected galaxy distribution of
  clusters have been presented by \citet{plionis_etal06}.}
\citet{buote_tsai95} were also the first to show that for the
simulated cluster they studied the shape of the X-ray isophotes
reflected the shape of the underlying three-dimensional gas
distribution and potential.

\citet{fang_etal09} compared the ellipticities of X-ray surface
brightness isophotes of clusters simulated with and without radiative
cooling and star formation. Focusing on a single cluster from the sample we
analyze in this paper, they showed that the shape of gas can be quite
flattened when gas cools significantly and settles into rotating thick
disk. They also showed that this flattening is detectable in the shape
of X-ray isophotes.  \citet{fang_etal09} also argued that the
flattened shape of the gas distribution in the simulated cluster
implies that gas does not trace potential in the inner regions due to
rotational support. Their results therefore demonstrate that shapes of
X-ray isophotes in cluster cores are a useful diagnostic of amount of
cooling and gas motions in cluster cores.  \citet{fang_etal09} have
also compared isophote shapes for synthetic {\it Chandra}
observations of a sample of clusters simulated with cooling to
observations and concluded that ellipticity profiles in runs with
radiative cooling do not match observations. They attributed the
discrepancy to ongoing, significant cooling in the cores of simulated
clusters that is absent from the cores of real clusters
\citep[e.g.,][]{peterson_fabian06}.

In this paper we present analysis of the three-dimensional shapes of
intracluster gas, DM, and underlying gravitational potential
using high-resolution cosmological simulations of galaxy clusters
formed in the $\Lambda$CDM cosmology. Our work extends the work of
\citet{fang_etal09} by presenting more detailed analysis of
three-dimensional shape profiles of gas, DM, and
gravitational potential for the full sample of clusters. In addition,
we focus on the effects of cumulative cooling and dissipation during
the entire cluster evolution on the shape of potential and gas
distribution at intermediate radii ($r>0.2r_{500}$), where dissipation
makes potential more spherical
\citep[e.g.,][]{kazantzidis_etal04}. This effect was not investigated
in \citet{fang_etal09}.  We show explicitly that dissipation leads to
more spherical shapes for ICM gas outside cluster cores ($0.1\lesssim
r/r_{500} \lesssim 1$)\footnote{Here and throughout this paper
  $r_{500}$ denotes the cluster-centric radius enclosing a mean
  overdensity of $500\rho_c(z)$, where $\rho_c(z)$ is the critical
  density of the universe at the redshift of analysis.}, reflecting
the corresponding effect on the DM distribution.  We also
find that the shape of gas matches the shape of the gravitational
potential at these radii in general, but deviates from it at smaller
radii and at $r \gtrsim r_{500}$ where assumption of the hydrostatic
equilibrium breaks down.  At smaller radii ($r\lesssim 0.1r_{500}$)
gas distributions in simulations with cooling become oblate,
reflecting the partially rotation-supported thick disks into which gas
settles as it loses its thermal energy by cooling, a result that is
qualitatively consistent with \citet{fang_etal09}.  We predict gas
shapes as may be determined observationally by estimating
ellipticities from mock X-ray maps of the same clusters and show that
one can constrain cluster gas physics by comparing ellipticity
profiles of simulations with and without dissipation to those of
observations.

This paper is organized as follows. In Section~\ref{sec:simulations}
we describe our cluster simulations.  In Section~\ref{sec:3d_results} we
describe the method of estimating axis ratios and the results for the
three dimensional shapes of clusters.  We give our estimates for ellipticities 
of mock X-ray maps as an observable prediction in
Section~\ref{sec:mock_xray}.  We provide a summary and discussion 
of our results in Section~\ref{sec:discussions}.

\section{The Simulations}
\label{sec:simulations}

\begin{table}[t]
\begin{center}
\caption{Properties of the Simulated Clusters at $z=0$}\label{tab:sim}
\begin{tabular}{l c c c c  }
\hline
\hline
Cluster ID\hspace*{5mm} & 
{$M_{500}$} & 
{$r_{500}$} & 
Relaxed (1)/Unrelaxed(0) \\
& {(10$^{14}$ $h^{-1} {M_{\odot}}$)}
& {($h^{-1}$ Mpc)} 
& $xyz$ \\
\hline
CL101 \dotfill & 9.02 & 1.16 & 000 \\
CL102 \dotfill & 5.45 & 0.98 & 000 \\
CL103 \dotfill & 5.70 & 0.99 & 000 \\
CL104 \dotfill & 5.40 & 0.98 & 111 \\
CL105 \dotfill & 4.86 & 0.94 & 001 \\
CL106 \dotfill & 3.47 & 0.84 & 000 \\
CL107 \dotfill & 2.57 & 0.76 & 100 \\
CL3 \dotfill & 2.09 & 0.71 & 111 \\
CL5 \dotfill & 1.31 & 0.61 & 111 \\
CL6 \dotfill & 1.68 & 0.66 & 000 \\
CL7 \dotfill & 1.42 & 0.63 & 111 \\
CL9  \dotfill & 0.83 & 0.52 & 000 \\
CL10 \dotfill & 0.67 & 0.49 & 111 \\
CL11 \dotfill & 0.90 & 0.54 & 000 \\
CL14 \dotfill & 0.77 & 0.51 & 111 \\
CL24 \dotfill & 0.35 & 0.39 & 010 \\
\hline
\end{tabular}
\end{center}
\end{table}

We analyze high-resolution cosmological simulations of 
16 cluster-sized systems in a flat {$\Lambda$}CDM model:
$\Omega_{\rm m}=1-\Omega_{\Lambda}=0.3$, $\Omega_{\rm b}=0.04286$,
$h=0.7$ and $\sigma_8=0.9$, where the Hubble constant is defined as
$100h{\ \rm km\ s^{-1}\ Mpc^{-1}}$, and $\sigma_8$ is the mass variance 
within spheres of radius $8h^{-1}$~Mpc and serves to normalize the 
power spectrum.  The simulations were performed using the 
Adaptive Refinement Tree $N$-body$+$gasdynamics code \citep{kra99,kra02}, 
an Eulerian code that uses adaptive refinement in space and time, and (non-adaptive)
refinement in mass \citep{klypin_etal01} to reach the high dynamic
ranges required to resolve the cores of halos formed in self-consistent
cosmological simulations. The simulations presented here are discussed
in detail in \citet{nagai_etal07a} and \citet{nagai_etal07} and we refer the reader
to these papers for more details. Here we summarize the relevant 
parameters of the simulations.

In order to assess the effects of gas cooling and star formation on
the cluster shapes, we conducted each cluster simulation with two
different prescriptions for gasdynamics. In one set of runs we treated 
only the standard gasdynamics for the baryonic component without 
either radiative cooling or star formation.  We refer to these as 
non-radiative (NR) runs.  In the second set of runs, we included gas 
cooling and star formation (CSF).  In the CSF runs, several physical
processes critical to various aspects of galaxy formation are
included: star formation, metal enrichment and thermal feedback due to
supernovae Type II and Type Ia, self-consistent advection of metals,
metallicity-dependent radiative cooling and UV heating due to a 
cosmological ionizing background \citep[see][for details
of the metallicity-dependent radiative cooling and star formation]{nagai_etal07}.
These simulations therefore follow the formation of galaxy clusters
starting from well-defined cosmological initial conditions and
capture the dynamics and properties of the intracluster medium (ICM) in a
realistic cosmological context.  However, some potentially relevant
physical processes, such as active galactic nuclei (AGNs) bubbles,
magnetic fields, and cosmic rays, are not included.  Consequently, 
the simulated cluster properties have limited application to real 
systems, most notably in the innermost cluster regions, 
where these processes are likely to be more important.

We ran our high-resolution simulations using a uniform 128$^3$ grid and
eight levels of mesh refinement in computational boxes of
$120\,h^{-1}$~Mpc comoving on a side for CL101--107 and
$80\,h^{-1}$~Mpc for CL3--24. These
simulations achieve a dynamic range of $32768$ and a formal peak
resolution of $\approx 3.66\,h^{-1}$~kpc and $2.44\,h^{-1}$~kpc,
corresponding to an actual resolution of $\approx 7\,h^{-1}$~kpc and
$5\,h^{-1}$~kpc, for the 120 and $80\,h^{-1}$~Mpc boxes, respectively.
Only regions of $\sim 3-10\,h^{-1}$~Mpc surrounding each cluster were
adaptively refined.  The remaining volume was followed on the uniform
$128^3$ grid.  The particle mass,  $m_{\rm p}$, corresponds to an
effective $512^3$ particles in the entire box, or a Nyquist wavelength
of $\lambda_{\rm Ny}=0.469\,h^{-1}$~Mpc and $0.312\,h^{-1}$~Mpc
comoving for CL101--107 and CL3--24, respectively.  These correspond
to $0.018\,h^{-1}$~Mpc and $0.006\,h^{-1}$~Mpc in physical units at
the initial redshifts of the simulations. The DM 
particle mass in the region around each cluster was $m_{\rm p}
\simeq 9.1\times 10^{8}\,h^{-1}\, {M_{\odot}}$ for CL101--107 and
$ m_{\rm p} \simeq 2.7\times 10^{8}\,h^{-1}\,{M_{\odot}}$ for
CL3--24, while other regions were simulated with lower mass
resolution. For this paper, we only report values at cluster-centric distances
larger than $0.03r_{500}$ where all clusters are well resolved. 

In order to test our simulation results against observations, we created mock {\it Chandra} 
X-ray images along three orthogonal projections for each simulated cluster. 
To minimize statistical fluctuation due to the Poisson noise, each image 
has an exposure time of 100 ks, corresponding to deep X-ray observations. 
Instrumental responses of {\it Chandra} were included in the mock image data.  
An overview of the methods used to generate the mock images is given in Section~\ref{sec:mock_xray}.  
Detailed descriptions of these mock images can be found in Section 3.1 of \cite{nagai_etal07}. 

To investigate the dependence of gas and halo shapes on the dynamical states of 
the clusters in our simulation set, we divided our sample into relaxed and 
unrelaxed clusters based on visual examinations of their mock X-ray images. 
A typical relaxed cluster is an object in which all three of its orthogonal 
images have a regular morphology and prominent substructures are absent.  
Details of the classification can be found in \cite{nagai_etal07a}.

In Table~\ref{tab:sim} we report $M_{500}$ (the mass within $r_{500}$), $r_{500}$, 
and our classification of each of the X-ray images along the 
three orthogonal projections as relaxed or unrelaxed for our sample of 
16 $z=0$ clusters. 

\section{Three Dimensional Shapes}
\label{sec:3d_results}

\begin{figure}[t]
\begin{center}
\includegraphics[scale=0.45]{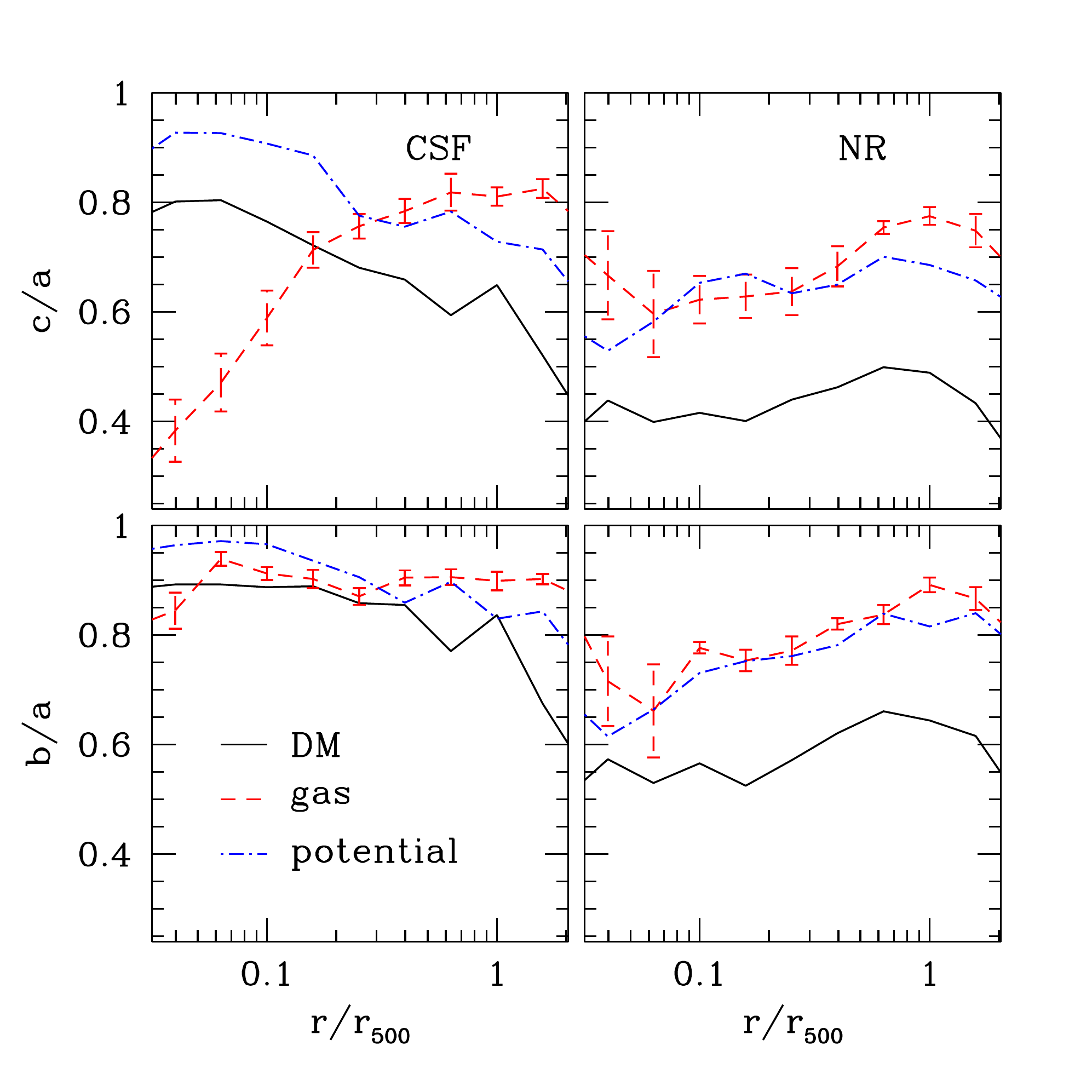}
\caption{
Average ellipsoidal axis ratio profiles for the {\em relaxed} $z=0$ clusters from the CSF run (left panels) 
and the NR run (right panels). The upper panels show
the profiles for the short-to-long axis ratio $c/a$, and the bottom panels show the profiles 
for the intermediate axis ratio $b/a$. 
In all panels, the solid line corresponds to dark matter (DM), the dashed line corresponds to gas, 
and the dot-dashed line corresponds to gravitational potential. The error bars show 1$\sigma$ 
error on the mean axis ratio for gas. The magnitude of the errors on the mean axis ratios for gas is similar
to those of DM and potential.
}
\label{fig:axis_pro}
\end{center}
\end{figure}

\begin{figure*}[htbp]
\begin{center}
\includegraphics[scale=0.8]{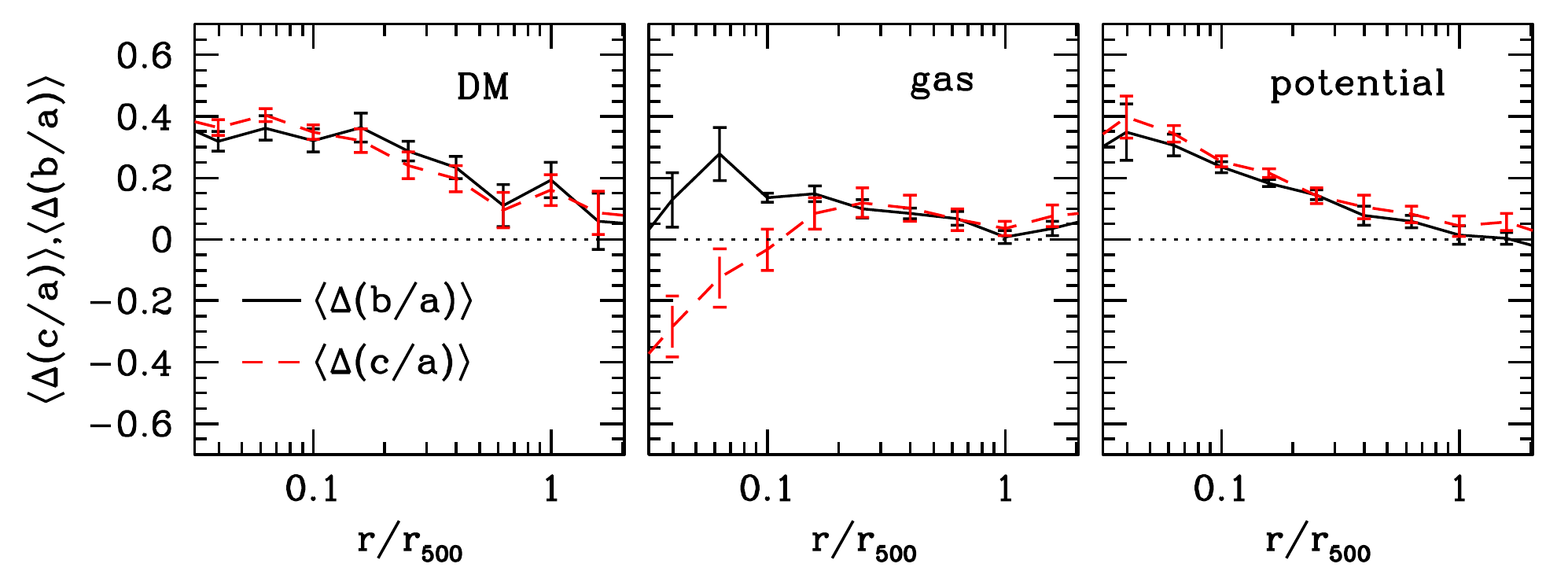}
\caption{
Difference in the axis ratio profiles of dark matter (left panel), 
gas (middle panel), and gravitational potential (right panel)
between CSF and NR runs averaged over relaxed clusters at $z=0$. 
The difference is defined as the CSF axis ratios minus the NR axis ratios. 
The black solid line is the difference for the short-to-long axis ratio, 
$\langle \Delta(c/a)\rangle$. 
The red dashed line is for the middle-to-long axis ratio, 
$\langle \Delta(b/a)\rangle$.}
\label{fig:axisdiff_pro}
\end{center}
\end{figure*}

\begin{figure*}[htbp]
\begin{center}
\includegraphics[scale=0.8]{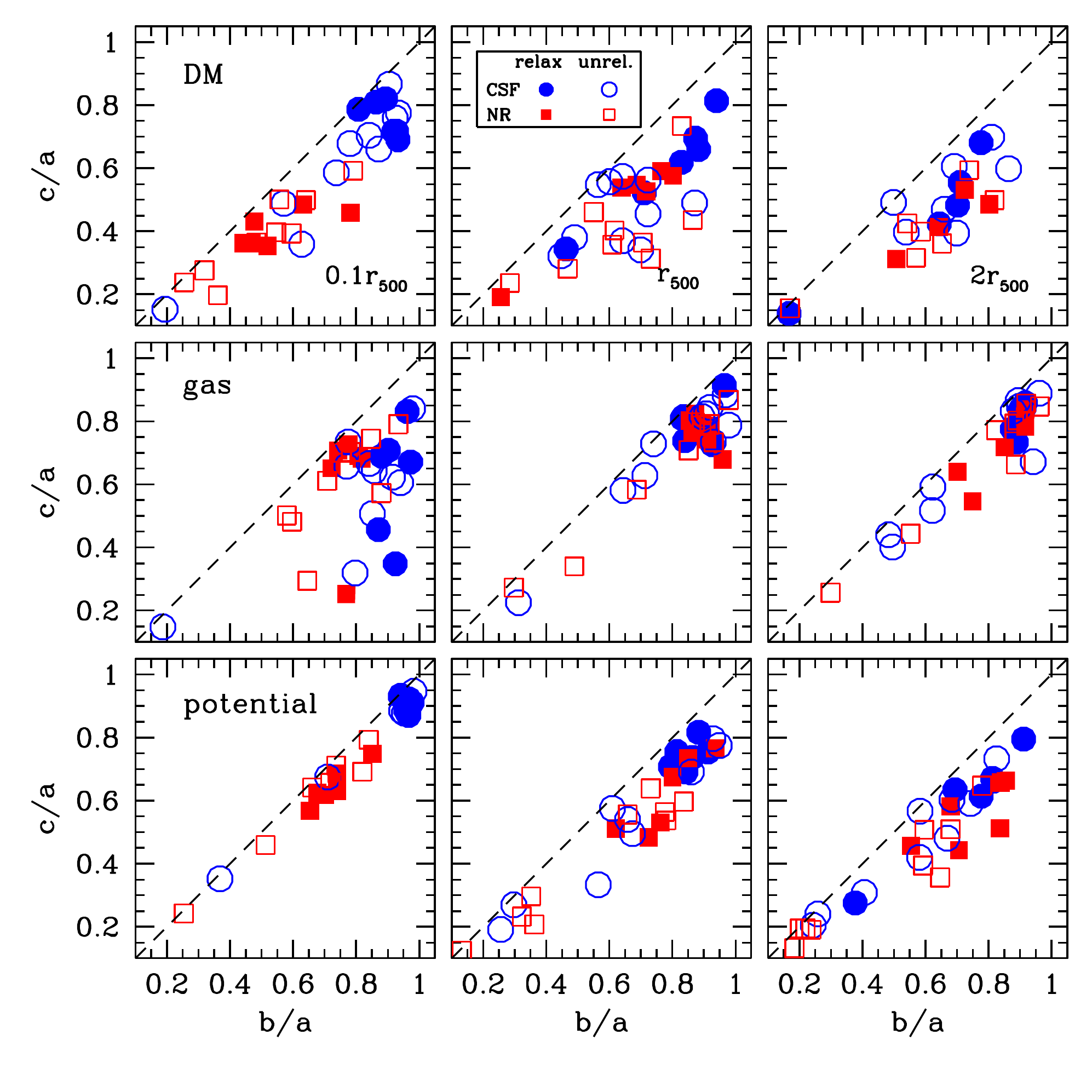}
\caption{
Minor-to-major axis ratio ($c/a$) vs. intermediate-to-major axis ratio ($b/a$) at 
$r=0.1r_{500}$ (left panels), $r=r_{500}$ (center panels) and $r=2 r_{500}$ (right panels) 
for DM (top panels), gas (middle panels), and gravitational potential (bottom panels) at $z=0$. 
Solid points are relaxed clusters and open points are unrelaxed clusters. 
CSF clusters are represented by circles, NR clusters are represented by squares. 
The dashed line is $c=b$. }
\label{fig:rqs_z0}
\end{center}
\end{figure*}

\begin{figure*}[htbp]
\begin{center}
\includegraphics[scale=0.8]{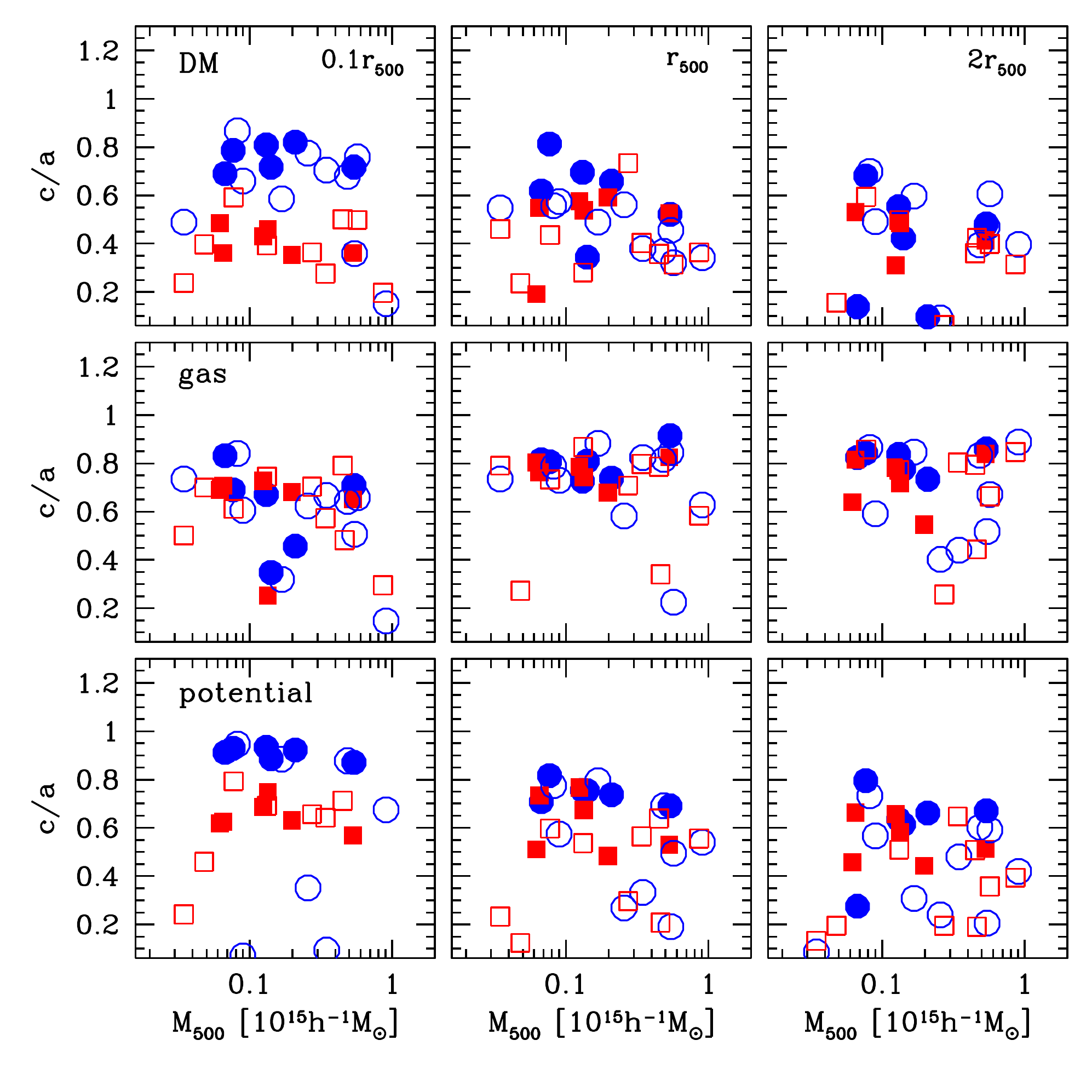}
\caption{
Minor-to-major axis ratio $c/a$ evaluated at $r=0.1r_{500}$ (left panels), $r_{500}$ (center panels),  
and 2$r_{500}$ (right panels) as a function of $M_{500}$.  DM shape profiles are shown in 
the top row of panels, gas in the middle row, and gravitational potential in the bottom row. 
All cluster profiles are computed at $z=0$.  
Solid points are relaxed clusters, and open points are unrelaxed clusters. 
CSF clusters are represented by circles, NR clusters are represented by squares.
}
\label{fig:sm_z0}
\end{center}
\end{figure*}

\subsection{Methods}
\label{subsection:methods}

We estimate axis ratios following \cite{dubinski_carlberg91} and \cite{kazantzidis_etal04}.  
For DM particles, we compute the inertia tensor
\begin{equation}
I_{ij} = \sum_{\alpha} {w_{\alpha} x_{\alpha,i} x_{\alpha,j}}, 
\label{eq:inertia_tensor}
\end{equation}
where $x_i$ is the $i$ coordinate of particle $\alpha$, 
$w_{\alpha}$ is the particle mass, 
and the sum is over all particles in a shell of width $\Delta r$. 
The principal axis lengths are obtained by diagonalizing $I_{ij}$ 
using an iterative scheme.  We begin by computing the inertia 
tensor in spherical shells and computing the axis ratios 
$q \equiv b/a$ and $s \equiv c/a<q$ using the eigenvalues of 
the inertia tensor (we adopt $a \ge b \ge c$ by convention).  
In the subsequent iterations, we compute principal axes at a given radius $r$
by summing over particles within ellipsoidal shells of width $\Delta r$, 
defined using orientations and values of the principal axes from
the previous iteration. The generalized ellipsoidal distance of particle $\alpha$ is  
\begin{equation}
r_{\alpha} = \sqrt{x^{\prime\ 2}_{\alpha}+\left(\frac{y^{\prime}_{\alpha}}{q}\right)^2+\left(\frac{z^{\prime}_{\alpha}}{s}\right)^2}, 
\label{eq:ralpha}
\end{equation}
where the primed coordinates are particle positions rotated into the frame of the principal 
axes of the inertia tensor.  The centers of all ellipsoids are fixed to be
the center of the cluster, defined as the location of the most bound
particle. This process is continued until convergence of $q$ and $s$
to better than $1\%$.

Note that \citet{dubinski_carlberg91} weighted each term in their calculation of the 
inertia tensor by $r_{\alpha}^{-2}$, to mitigate the influence of substructures, 
prevalent at the outskirts of halos, on axis ratios computed within a certain radius.  
As in \cite{kazantzidis_etal04}, we find that this weighting makes only a small difference 
for axis ratios computed within narrow radial bins so we do not employ this weighting in 
our analysis. In this paper we use the mean $r_{\alpha}$ (Equation~(\ref{eq:ralpha}))of particles 
within each ellipsoidal shell (equivalent to the length of the major axis of the shell) as our 
measure of cluster-centric distance, unless stated otherwise. 

We estimate the axis ratios for gas in a similar way with weights 
$w_{\alpha} = \rho_{{\rm gas},\alpha}V_\alpha$, where $\rho_{{\rm gas},\alpha}$ and $V_{\alpha}$ 
are the gas density and volume of the $\alpha$-th grid cell.  
We estimate the axis ratios of the surfaces of constant gravitational potential 
by computing the inertia tensor of all cells with potential within some range 
$[\Phi,\Phi+\Delta \Phi]$, taking $w_{\alpha}=1$ for all such cells. 

Large subhalos can bias the axis ratios in a given radial bin to
lower values, generating a local fluctuation in the axis ratio
profile. To minimize fluctuations due to substructures we
remove particles bound to subhalos of masses $M_{\mathrm{sub}} > 10^{12} \, h^{-1}M_{\odot}$. 
The identification of subhalos and bound particles follows 
the procedure described in \citet{kravtsov_etal04}.  

\subsection{Results}

Figure~\ref{fig:axis_pro} shows the axis ratio profiles for dark
matter, gas, and gravitational potential averaged over the relaxed
clusters at $z=0$ for the CSF and NR runs respectively.  Results are
similar for unrelaxed clusters but with considerable scatter. In both
CSF and NR runs, gravitational potential is much more spherical than
DM.  The potential of a given thin triaxial mass shell with
axis ratios $q$ and $s$ (the homeoid) is constant within the shell and
has a triaxial shape outside of isopotential surfaces defined by the
ellipsoid \citep[e.g.,][]{binney_tremaine08}:
$$r^2=\frac{x^2}{\tau+1}+\frac{y^2}{\tau+q^2}+\frac{z^2}{\tau+s^2},$$
where $\tau$ is the label of the surface. The shape of the
isopotential surfaces corresponding to a given ellipsoidal shell
therefore becomes more spherical as the distance from the shell
increases. Total potential at a given distance from the cluster is the
superposition of the potentials generated by ellipsoidal shells within
this distance and a constant potential generated by ellipsoidal shells
outside. The shape of the isopotential surfaces therefore will always
be more spherical than the shape of the underlying mass distribution
that gives rise to the potential.

Figure~\ref{fig:axis_pro} shows that at $r \gtrsim 0.1r_{500}$ the
shapes of DM, gravitational potential, and gas are more
spherical in the relaxed CSF clusters than in the relaxed NR clusters,
an effect identified previously in several studies (see
Section~\ref{section:intro}). This effect appears to be due to the
fact that potential becomes more spherical at large radii for a more
concentrated mass distribution resulting from baryon dissipation and
adiabatic response of the particle orbits to such change of potential
\citep{debattista_etal08,valluri_etal10}.

The effect is more apparent in
Figure~\ref{fig:axisdiff_pro}, which shows the differences in axis
ratios between the relaxed CSF and NR clusters defined as
\begin{equation}
\Delta(c/a) = \left(c/a\right)_{\rm CSF}-\left(c/a\right)_{\rm NR}
\end{equation}
and similarly for $b/a$.  It is evident that baryonic dissipation causes relaxed
DM halos to become significantly more spherical in
their inner regions, an effect that remains significant out to  
$r_{500}$ and beyond.  
The average axis ratio shifts drop from 
$\langle \Delta(c/a)\rangle \sim \langle \Delta(b/a)\rangle \sim 0.3$ at $0.1 r_{500}$ to 
$\langle \Delta(c/a)\rangle \sim \langle \Delta(c/a)\rangle \sim 0.1$ at $r_{500}$.  
Changes in $c/a$ and $b/a$ are very nearly the same at all radii.  

For gas, the average change in both axis ratios is
$\langle \Delta(c/a)\rangle \approx \langle \Delta(b/a)\rangle \approx
0.1$ at $0.2r_{500}$, and is decreasing slowly to zero at $r_{500}$.
However, at smaller radii, $r \lesssim 0.1r_{500}$, there is a
positive change in the intermediate axis ratio $b/a$ and a
negative change in the short-to-long axis ratio $c/a$.  At $r\approx
0.05r_{500}$, $\langle \Delta(c/a)\rangle \approx -0.3$ and continues
to decrease with radius, while $\langle \Delta(b/a)\rangle$ increases
to $\approx 0.2$ at $0.06r_{500}$ and decreases inward.  The
significant decrease in $c/a$ indicates that the gas assumes an oblate
shape in the inner regions of the relaxed CSF clusters compared to the
prolate gas shapes in their NR counterparts.  This is consistent with
the expectation that ongoing cooling in the inner cluster region leads
to formation of a thick oblate disk \citep{fang_etal09}.

The difference between the shapes of the gravitational potential of
relaxed CSF and NR clusters is qualitatively similar to that of the
DM. The average change in axis ratios is
$\langle \Delta(c/a)\rangle \approx \langle \Delta(b/a)\rangle \approx
0.2$ at $r\approx 0.1r_{500}$, and decreases to nearly zero at
$r_{500}$.  There is little difference between
$\langle \Delta(c/a)\rangle$ and $\langle \Delta(b/a)\rangle$.

Finally, Figure~\ref{fig:axis_pro} shows that gas traces the
shape of gravitational potential for relaxed NR clusters in the radial range
$0.06\lesssim r/r_{500}\lesssim 1$, indicating that gas is in
approximate hydrostatic equilibrium within the potential. 
However, in the relaxed CSF clusters the gas and potential shapes only match at
over a relatively narrow range of radii, $0.2\lesssim r/r_{500}\lesssim 0.4$, 
signaling departures from hydrostatic equilibrium.  
We discuss this further in Section~\ref{subsec:gas_vs_pot}.

\begin{table*}[t]
\begin{center}
\caption{axis ratio $c/a$ of the 16 simulated clusters}\label{tab:axis_ratio}
\begin{tabular}{ c c c | c c c c | c c c c }
\hline
\hline
\multicolumn{3}{c}{}&\multicolumn{4}{|c}{$z=0$} &\multicolumn{4}{|c}{$z=0.6$}\\
\hline
&&$r/r_{500}=$&0.1&0.5&1.0&2.0&0.1&0.5&1.0&2.0\\
\hline
All&DM&CSF&$0.66\pm0.05$&$0.47\pm0.05$&$0.54\pm0.04$&$0.39\pm0.06$&
$0.59\pm0.06$&$0.32\pm0.06$&$0.33\pm0.07$&$0.36\pm0.06$\\
&&NR&$0.37\pm0.03$&$0.38\pm0.04$&$0.41\pm0.05$&$0.28\pm0.05$&
$0.40\pm0.04$&$0.38\pm0.04$&$0.41\pm0.04$&$0.14\pm0.05$\\
\cline{2-11}
&potential&CSF&$0.59\pm0.10$&$0.58\pm0.08$&$0.55\pm0.06$&$0.53\pm0.05$&
$0.59\pm0.10$&$0.50\pm0.07$&$0.50\pm0.07$&$0.38\pm0.05$\\
&&NR&$0.49\pm0.08$&$0.57\pm0.07$&$0.56\pm0.05$&$0.49\pm0.05$&
$0.33\pm0.08$&$0.44\pm0.06$&$0.54\pm0.06$&$0.22\pm0.07$\\
\cline{2-11}
&gas&CSF&$0.58\pm0.05$&$0.67\pm0.67$&$0.74\pm0.04$&$0.70\pm0.06$&
$0.68\pm0.04$&$0.69\pm0.05$&$0.69\pm0.06$&$0.59\pm0.05$\\
&&NR&$0.58\pm0.05$&$0.65\pm0.05$&$0.67\pm0.06$&$0.63\pm0.06$&
$0.39\pm0.07$&$0.55\pm0.06$&$0.66\pm0.04$&$0.26\pm0.08$\\
\hline
\hline
Relaxed&DM&CSF&$0.76\pm0.02$&$0.63 \pm0.05$&$0.67\pm0.04$&$0.46\pm0.10$&
$0.51\pm0.12$&$0.39\pm0.10$&$0.40\pm0.09$&$0.39\pm0.13$\\
&&NR&$0.41\pm0.02$&$0.46\pm0.06$&$0.48\pm0.07$&$0.30\pm0.10$&
$0.32\pm0.04$&$0.42\pm0.03$&$0.46\pm0.01$&$0.17\pm0.11$\\
\cline{2-11}
&potential&CSF&$0.91\pm0.01$&$0.82\pm0.02$&$0.71\pm0.02$&$0.65\pm0.03$&
$0.82\pm0.02$&$0.70\pm0.02$&$0.69\pm0.03$&$0.49\pm0.07$\\
&&NR&$0.65\pm0.03$&$0.72\pm0.06$&$0.67\pm0.05$&$0.58\pm0.02$&
$0.36\pm0.12$&$0.64\pm0.02$&$0.69\pm0.04$&$0.22\pm0.13$\\
\cline{2-11}
&gas&CSF&$0.61\pm0.07$&$0.79\pm0.05$&$0.80\pm0.03$&$0.82\pm0.02$&
$0.68\pm0.06$&$0.78\pm0.02$&$0.80\pm0.04$&$0.75\pm0.05$\\
&&NR&$0.61\pm0.07$&$0.74\pm0.03$&$0.76\pm0.02$&$0.72\pm0.05$&
$0.43\pm0.13$&$0.67\pm0.04$&$0.72\pm0.05$&$0.28\pm0.18$\\
\hline
\end{tabular}
\end{center}
\end{table*}

Figure~\ref{fig:rqs_z0} shows $c/a$ versus $b/a$ at $r/r_{500} = (0.1,1.0,2.0)$
for our relaxed and unrelaxed cluster samples. The average $c/a$ values are summarized in
Table~\ref{tab:axis_ratio}.  As could be expected, the DM distribution in
unrelaxed clusters is more triaxial on average compared to relaxed clusters. 
The DM halos in the NR clusters become more triaxial at smaller cluster-centric radii, 
consistent with previous findings in dissipationless simulations 
\citep[e.g.,][]{allgood_etal06}.  Conversely, the CSF clusters are rounder 
at small radii due to the effects of baryonic dissipation on halo shapes which 
are most prominent near halo centers \citep[e.g.][]{kazantzidis_etal04}.

Intracluster gas is very spherical at large radii in almost all
relaxed NR and CSF clusters.  
Unrelaxed clusters tend to have more triaxial gas distribution compared to relaxed clusters.  
A similar trend can be seen for potential. 

Figure~\ref{fig:sm_z0} shows $c/a$ as a function of cluster mass
$M_{500}$.  There is at most only a weak trend of decreasing $c/a$
with cluster mass for both the CSF and NR relaxed clusters seen in DM and
potential. This is consistent with the results of studies based on large statistical
samples of halos in dissipationless simulations 
\citep{kasun_evrard05,allgood_etal06,gottloeber_yepes07,ragonefigueroa_plionis07,maccio_etal08} which 
find $c/a\propto M^{-[0.03-0.05]}$.  Our sample of clusters is too small to detect
such a trend.

\begin{figure}[htbp]
\centering
\includegraphics[scale=0.44]{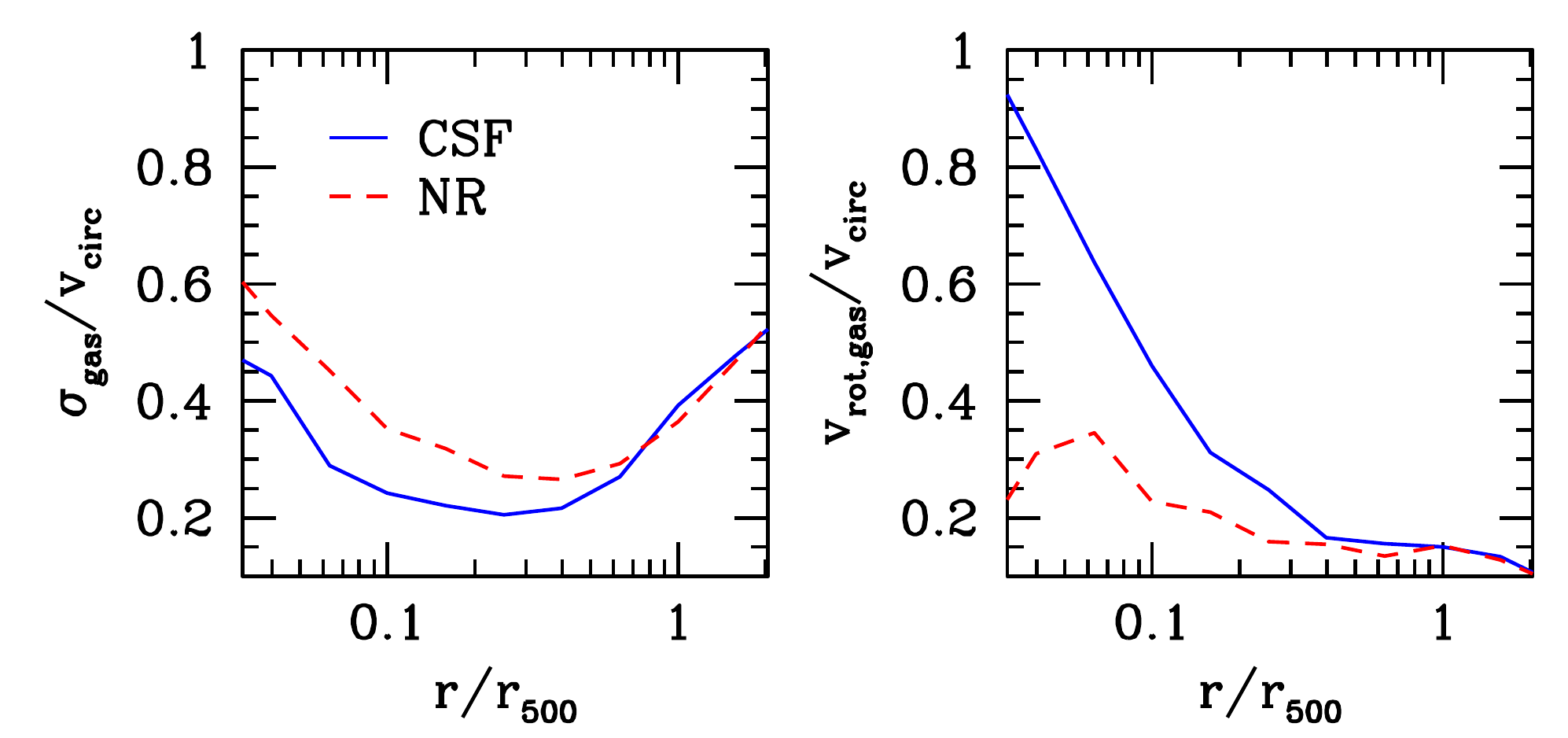}
\caption{
Relative contributions of random and rotational motions supporting gas against 
gravity in the relaxed CSF and NR clusters.  The left panel shows the ratio of the gas velocity 
dispersion to the circular velocity as a function of cluster-centric position.  The 
right panel shows the ratio of the rotational velocity to the circular velocity. 
}
\label{fig:sigma_pro}
\end{figure}

\subsection{Comparing the Shape of Intracluster Gas to the Shape of Gravitational Potential}
\label{subsec:gas_vs_pot}

In hydrostatic equilibrium, the isodensity surfaces of gas should 
trace the isopotential surfaces.  Consequently, any differences between 
the measured shapes of gas and potential indicate deviations from 
hydrostatic equilibrium.  Figure~\ref{fig:axis_pro} shows that 
such differences are present at small radii $r\lesssim 0.06 r_{500}$ in relaxed NR 
and $r\lesssim 0.2 r_{500}$ in relaxed CSF clusters, 
and at larger radii $r\gtrsim r_{500}$ and $\gtrsim 0.6r_{500}$ in the NR and CSF runs, 
respectively. All of these differences are due to the presence of gas bulk motions, 
although the nature and origin of these motions is different at small and large radii 
and in NR and CSF runs \citep[see][]{lau_etal09}. 

Figure~\ref{fig:sigma_pro} shows
the ratio of the isotropic gas velocity dispersion $\sigma_{\rm gas}$ and gas rotational velocity
(calculated within each ellipsoidal gas shell) to the circular
velocity (defined as $v_{\rm circ} \equiv\sqrt{GM(<r)/r}$, where $r$
is the mean ellipsoidal radius of the shell (Equation~(\ref{eq:ralpha})).  
This ratio indicates the relative contribution of
random gas motions in support against gravity.  For gas in perfect
hydrostatic equilibrium $\sigma_{\rm gas}$ should be zero.  However,
we see that in both the NR and CSF clusters 
$\sigma_{\rm gas} \approx 0.2-0.4v_{\rm circ}$ for 
$0.1\lesssim r/r_{500}\lesssim 1$,
corresponding to a fraction of pressure support due to random gas
motion of about 16\%.  These motions are dynamical in origin (due to
accretion of gas and motions of cluster galaxies) and are not affected
by cooling. 

The gas motions do not affect the shape of the gas
significantly at radii where $\sigma_{\rm gas}/v_{\rm circ}\lesssim
0.4$, but have significant effect for larger values of 
$\sigma_{\rm gas}/v_{\rm circ}$, as is apparent from the differences between the
shapes of gas and potential at small and large radii in the NR
clusters. The net effect of these motions is relatively rounder gas distributions 
compared to the shapes of isopotential surfaces. 

In addition to random motions, the gas in the CSF runs exhibits 
significant ordered rotational motions at $r \lesssim 0.3r_{500}$.  
These motions arise due to the angular momentum of the
ICM gas, which leads to rotation as gas cools and contracts. These
motions are responsible for the deviations between the shapes of gas
and potential at these radii. As we can see in 
Figure~\ref{fig:axis_pro}, the effect of the ordered motions is 
opposite to that of random motions, namely, rotational motions lead to 
gas distributions that are more flattened compared to 
isopotential surfaces. 

The effect of cooling, and the ordered rotational motions that result
from cooling, manifests as a rapid decrease of $c/a$ at small radii,
even as $b/a$ remains approximately constant. The effect of random gas
motions results in a rapid increase of both the $c/a$ and $b/a$
ratios. Measurements of the ellipticity profiles in clusters can
therefore constrain the cooling of gas and magnitude of residual gas
motions in their cores. Although X-ray spectroscopy is a much more
sensitive tool to constrain the contemporary cooling, rotational
motions and their effect on gas ellipticity also constrain the net
cooling that has been occurred in the past.  They can therefore
provide potentially useful and complementary constraints on the
thermal history of gas in cluster cores over the past several billion
years. However, in order to make sure that such constraints are
feasible we must check that the trends observed in the
three-dimensional gas distribution are evident in the X-ray photon maps
of these clusters.  We present such an analysis in the next section.

\section{Ellipticity of X-ray images of simulated clusters}
\label{sec:mock_xray}

\subsection{Methods}
\label{sub:mock_xray_methods}

Although one can expect that at radii where gas traces gravitational
potential the X-ray isophotes can be used to quantify its shape, it is
not immediately clear whether this can be done with the accuracy
sufficient to detect the effects of cooling discussed in the previous
section.  In order to compare our results on intracluster gas shapes
in Section~\ref{sec:3d_results} to observations, we estimate the
ellipticities of mock X-ray photon maps generated from the same $z=0$
simulated clusters.  Below we give a brief overview of the mock X-ray
maps.  We refer the reader to Section~3.1 of \cite{nagai_etal07} for a
detailed description.

First we create X-ray flux maps of the simulated clusters viewed along
three orthogonal projections. The flux map is computed by projecting
the X-ray emission of hydrodynamic cells enclosed within $3r_{\rm vir}$ of
a cluster along the line of sight.  The X-ray emissivity in each
computational grid cell is computed as a function of proton and
electron densities, gas temperature and metal abundance. Emission from
gas with temperature less than $10^5$~K is excluded as it is below the
{\it Chandra} bandpass.  We then convolve the emission spectrum with
the response of the {\it Chandra} front-illuminated CCDs and draw a
number of photons at each position and spectral channel from the
corresponding Poisson distribution.  Each map has an exposure time of
100 ks (typical for deep observations) and includes a background with
the intensity corresponding to the quiescent background level in the
ACIS-I observations \citep{markevitch_etal03}.  The resolution of all
the maps is 6 kpc pixel$^{-1}$. We use at least 25 pixels per bin for
ellipticity measurements.

From these data, we generate images in the 0.7--2 keV 
band and use them to identify and mask out all detectable 
small-scale clumps, as is routinely done in observational studies. 
Our clump detection is fully automatic and based on the wavelet 
decomposition algorithm \citep{vikhlinin_etal98}. 
The holes left by masking out substructures in the photon map are filled in by the values 
from the decomposed map of the largest scale in wavelet analysis.  
We have tested that this method preserves the global shape of the photon distribution well. 
The background is removed when estimating ellipticities as it 
can bias them low at radii where background dominates the intrinsic emission. 
Throughout this paper we assume the cluster redshift is $z_{obs}=0.06$ for the $z=0$ sample. 

\begin{figure*}[htbp]
\begin{center}
\includegraphics[scale=0.4]{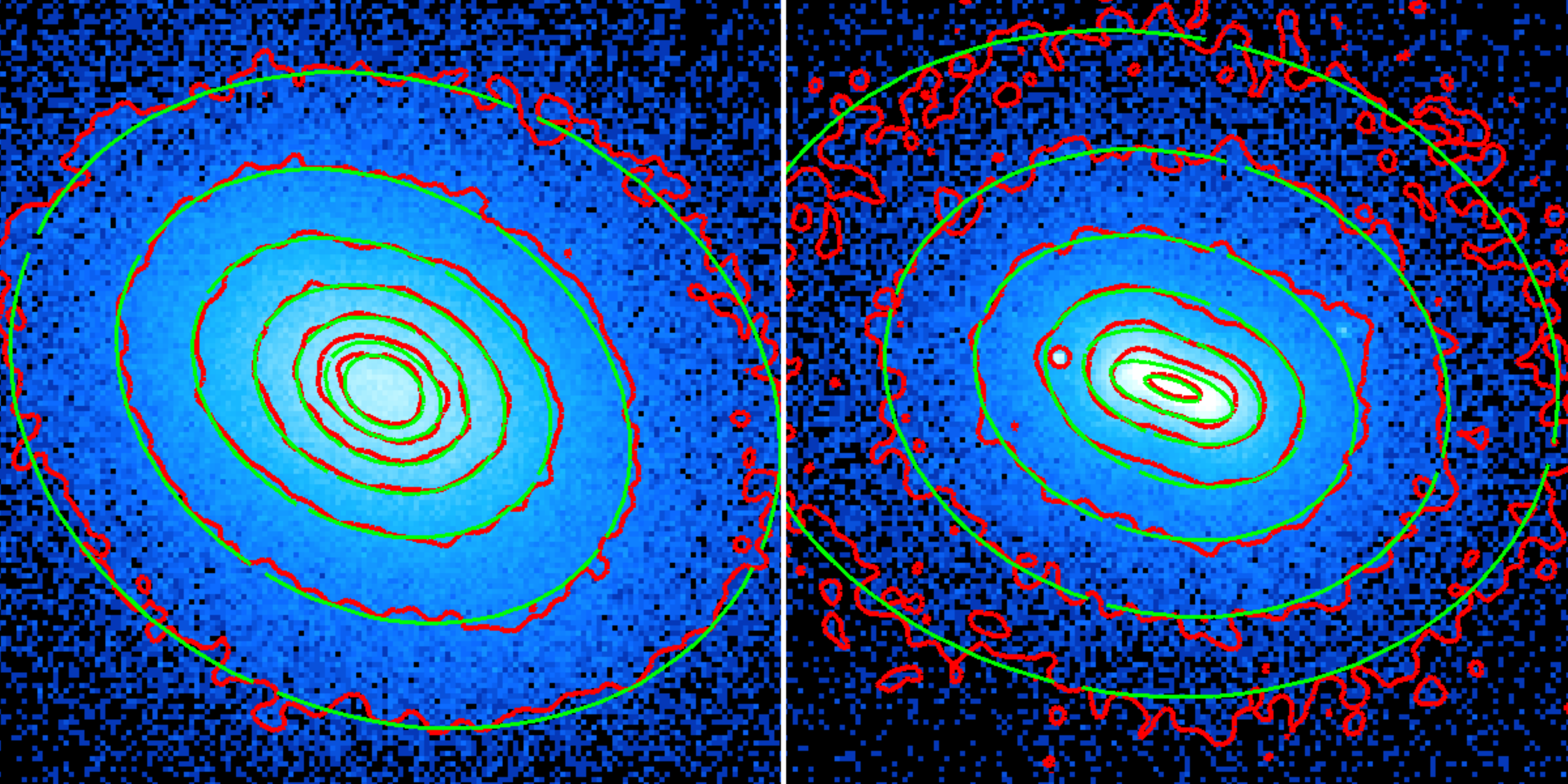}
\caption{
X-ray photon maps for the $y$-projection of CL7 in the NR run (left panel) and 
the CSF run (right panel). T
he length of the side of each panel is $0.9$ Mpc ($r_{500} = 0.854$ Mpc for the NR run, 
and $r_{500} = 0.891$ Mpc for the CSF run). 
Also shown are the isophotal contours 
(red) and the best-fit ellipses (green) using the method described in 
Section~\ref{sub:mock_xray_methods}.       
}
\label{fig:CL7y}
\end{center}
\end{figure*}

We define the ellipticity as
\begin{equation}
\epsilon \equiv 1-\frac{b}{a},
\label{eq:ellipticity}
\end{equation}
where $a$ and $b$ are the semi-major and the semi-minor axes of the
projected ellipse respectively.  The ellipticities of the X-ray photon
distributions are determined using the same algorithm, based on the
inertia tensor, as was used for the three-dimensional shapes in
Section~\ref{sec:3d_results}.  Instead of using particle mass, we
calculate the inertia tensor using weights given by the photon counts
in each map pixel. We have estimated the ellipticity within both
differential radial bins, and cumulative ellipticity within a given
radius. In addition we have estimated ellipticity within radial shells
defined by isophotes with different flux levels (isophotal
bins). We find from visual comparison of the X-ray contours and the
fitted ellipses that radial bins give reliable ellipsoidal fits to the X-ray isophotes 
(see Fig.~\ref{fig:CL7y} for an example). 

\begin{figure}[htbp]
\begin{center}
\includegraphics[scale=0.44]{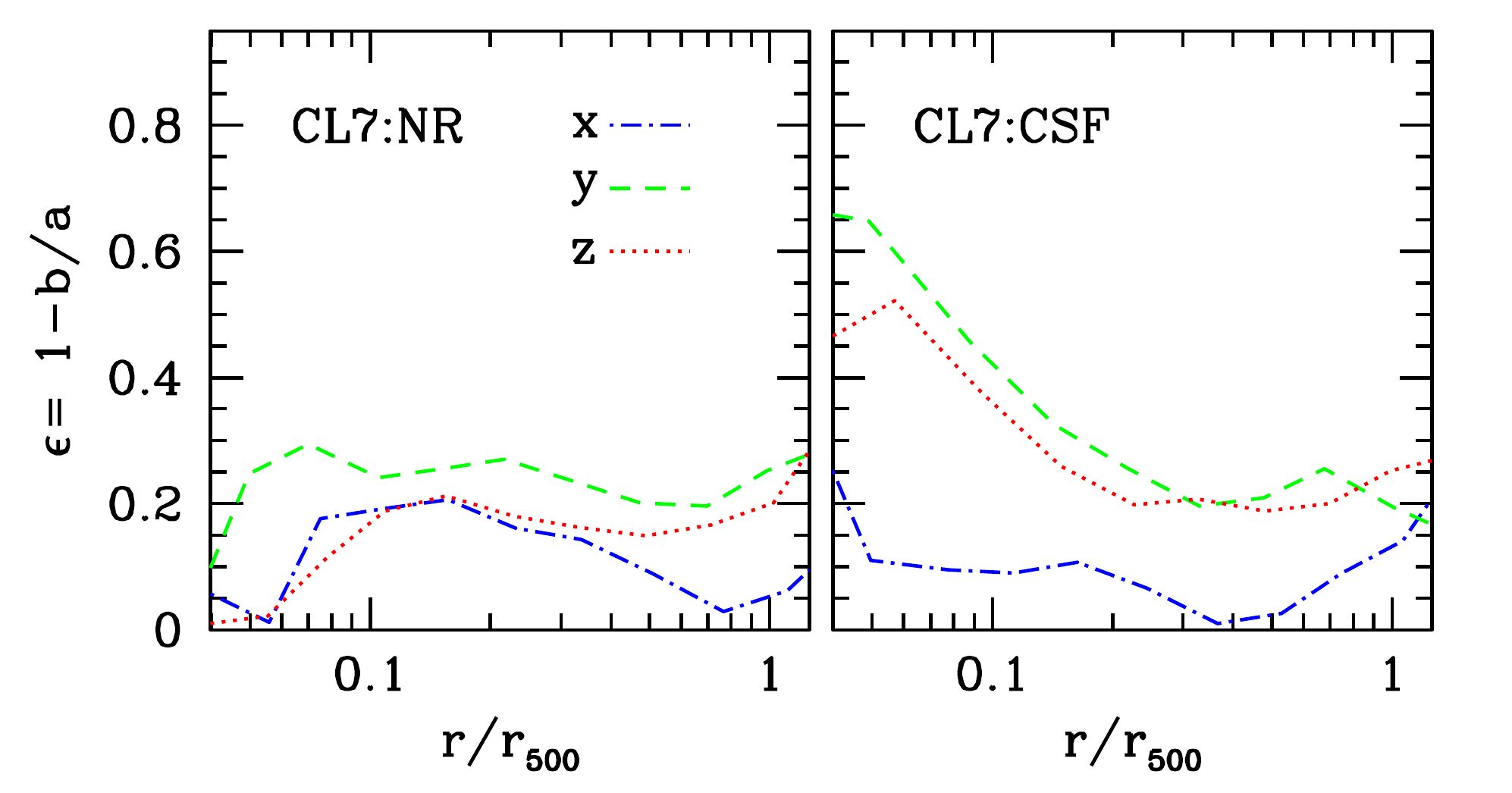}
\caption{
Ellipticity profiles of the CL7 cluster with NR (left panel, CL7:NR)
and in the CSF run (right panel, CL7:CSF) viewed along three 
orthogonal projections ($x$, $y$, $z$).
}
\label{fig:CL7_pro}
\end{center}
\end{figure}

\subsection{Test Case: CL7}
\label{sub:CL7}

We test our method for estimating ellipticity using CL7, one of the most
relaxed clusters in our sample. Figure~\ref{fig:CL7y} shows the 
mock X-ray maps for the $y$-projection of the NR and CSF runs 
(labeled as CL7:NR and CL7:CSF respectively).  Also shown are the isophotal contour 
lines (red) and the fitted ellipses of the photon distribution (green) 
derived using the method described in Section~\ref{sub:mock_xray_methods}. 
  
These images show that the ellipticities in the outer regions are similar for the 
NR and CSF runs, while they are very different in the core ($\lesssim 0.1 r_{500}$).  
The ellipticity increases towards small radii for the CSF run, while the opposite trend 
is seen for the NR run, where the isophotes are becoming slightly more spherical in the 
inner regions.  These images also demonstrate that the best-fit ellipse (indicated with 
green contour) describes the actual photon distribution (indicated with red
contours) well at all radii, including the flattened gas disk in the inner most regions 
of the CSF run.

Figure~\ref{fig:CL7_pro} shows the ellipticity profiles of the CL7:NR and CL7:CSF runs viewed 
along three orthogonal projections. 
Note that we have scaled the radius to $r=\sqrt{ab}$ (where $a$ and $b$ are the semi-major and 
semi-minor axes of the fitted ellipse), to be consistent with \cite{fang_etal09}.  
In the NR run, the ellipticity profiles for 
all three projections are quite similar over the range $r \gtrsim 0.1 r_{500}$, 
with nearly constant ellipticities of $\epsilon \approx 0.2$ (though it decreases to $\epsilon \approx 0.05$ 
at $r_{500}$ in the $x$-projection).  For the CSF run, two projections have very similar ellipticity 
profiles (the $y$- and $z$-projections), with $\epsilon \approx 0.2$ at $r \approx r_{500}$ and increasing 
towards smaller radii, reaching $\epsilon \approx 0.4-0.45$ at $r \approx 0.1r_{500}$.  
The $x$-projection, on the other hand, exhibits low ellipticity comparable to the NR run, $\epsilon \approx 0.1$,
and shows no strong trend with radius. These results are consistent with the picture that the inner
regions of the CSF run consist of a disk-like structure in the core, which appears highly 
elliptical when viewed edge-on (the $y$- and $z$-projections), while the flattened disk
appears more spherical when viewed face-on (in the $x$-projection).

Recently, the same set of simulated clusters are analyzed \citep[hereafter F09]{fang_etal09}. 
In particular, the ellipticity profiles of CL7 was presented in detail
(cf. Figure~7 in F09).  Although our results are in
qualitative agreement with those of F09 as to the 
formation of flattened gas structure in the cluster core, we find
substantial quantitative differences regarding the impact of the
baryonic dissipation.  For example, we find that the ellipticity
profiles for the CL7:CSF run are generally lower than the
corresponding F09 profiles.  In the $y$-projection at
$r=0.3r_{500}$, we find $\epsilon \approx 0.2$ as opposed to
the F09 value of $\epsilon \approx 0.5$.  Our
ellipticities are even significantly (by $\approx 0.2$-$0.3$) smaller in
the cluster core. In addition, Figure~\ref{fig:CL7_pro} shows that
significant flattening of the isophotes due to cooling is confined to
$r\lesssim 0.2r_{500}$, not out to $0.4r_{500}$ as stated
by F09.

We have also checked ellipticities from cumulative bins rather than annular bins 
and found that the ellipticity increases only slightly to $\epsilon = 0.3$ at $r=0.3r_{500}$.  
By inspection, the F09 surface brightness map for the $y$-projection 
of CL7:CSF, shown in their Figure~1, appears to be inconsistent with 
$\epsilon = 0.5$ at $r=0.3r_{500}$, but is consistent with the isophotes 
and our fits shown in Figures~\ref{fig:CL7y} and \ref{fig:CL7_pro}. 
To pin down the discrepancy, we focus on the $y$-projection of CL7:CSF and 
use our ellipticity code on both our FITS file and the FITS file given by the authors of F09. 
For consistency, we use cumulative bins in both cases. We find that 
the ellipticity profiles derived from the two different FITS files have very similar shape. 
The only difference is that the ellipticity profile from their FITS file appears to be shifted
to larger radii by a factor of 1.75. Upon inspection of the two FITS file, we find that the
length scale of their FITS file is 1.75 larger compared to our own FITS file. The origin of this
1.75 factor is unexplained in F09. 
We further find that the ellipticity profile estimated using our code on 
their FITS file is nearly identical to the profile for the $y$-projection shown in Figure~7 of F09. 
Thus we conclude that our discrepancy with F09 is most likely due to this 1.75 factor in their
FITS files. 

Therefore, although we see effects qualitatively similar to those pointed out 
by F09, the actual magnitude of the effect of dissipation of ellipticity 
of the ICM gas is much smaller than measured by F09 and is confined 
to the inner $r\lesssim 0.2r_{500}$ of the clusters. 

\begin{figure*}[htbp]
\begin{center}
\includegraphics[scale=0.8]{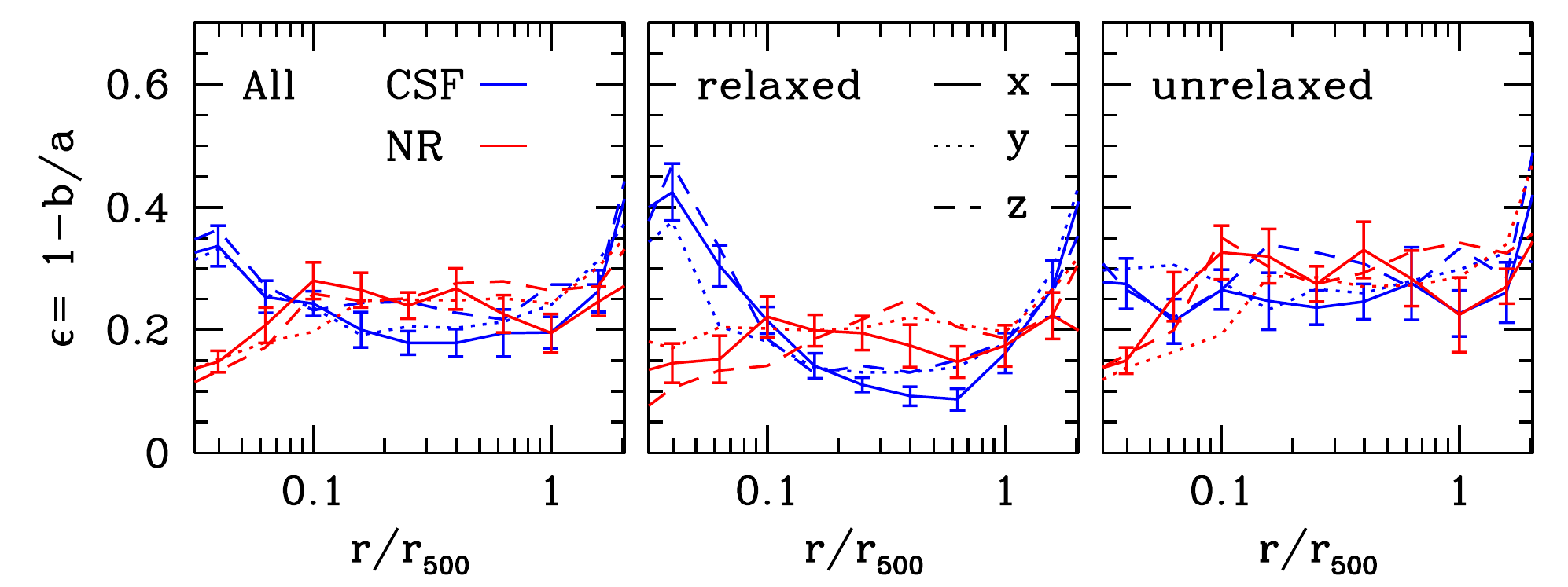}
\caption{
Ellipticity profiles averaged over mock X-ray maps of simulated clusters in the $x$-, $y$-, and $z$-projections (solid, dotted, and dashed lines, respectively). 
The blue lines represent the profiles for the CSF simulations, and the red lines for the NR runs.
The left panel shows the profiles averaged over all clusters, 
the middle panel shows the profiles averaged over 21 projections, in which clusters 
appear relaxed, and the right panel shows ellipticity profiles averaged over projections
in which clusters appear unrelaxed.  The errorbars show the
$1\sigma$ errors of the mean values for the ellipticity profile for the $x$-projections. 
The errors are similar for the profiles in the $y$- and $z$-projections. }
\vspace{1.5mm}
\label{fig:ellip_pro}
\end{center}
\end{figure*}

\begin{table*}[htbp]
\begin{center}
\caption{Ellipticity of the 48 mock x-ray maps for the $z=0$ clusters}\label{tab:ellipticity}
\begin{tabular}{ c c | c c c c }
\hline
\hline
&$r/r_{500}=$&0.05&0.3&1.0&2.0\\
\hline
All&CSF&$0.27\pm0.03$&$0.20\pm0.02$&$0.17\pm0.02$&$0.40\pm0.04$\\
&NR&$0.19\pm0.03$&$0.30\pm0.03$&$0.20\pm0.02$&$0.38\pm0.03$\\
\hline
Relaxed&CSF&$0.32\pm0.06$&$0.12\pm0.01$&$0.14\pm0.02$&$0.32\pm0.06$\\
&NR&$0.17\pm0.05$&$0.21\pm0.02$&$0.13\pm0.01$&$0.31\pm0.04$\\
\hline
Unrelaxed&CSF&$0.25\pm0.03$&$0.27\pm0.04$&$0.20\pm0.03$&$0.43\pm0.05$\\
&NR&$0.21\pm0.04$&$0.37\pm0.06$&$0.25\pm0.03$&$0.45\pm0.04$\\
\hline
\end{tabular}
\end{center}
\end{table*}

\subsection{Results}
\label{sub:results}
Figure~\ref{fig:ellip_pro} shows the ellipticity profiles derived from 
mock X-ray maps averaged over the three orthogonal projections separately
for all clusters as well as subsets of the images of relaxed clusters and 
the images of unrelaxed clusters (see Table~\ref{tab:sim}). 
The radial coordinate here is actually the semi-major axis $a$
of the fitted ellipse in units of $r_{500}$.  We note explicitly 
that the three-dimensional results in Section~\ref{sec:3d_results} 
were quoted as axis ratios, while the results of this section 
are given in ellipticities ($\epsilon = 1 - b/a$). 

The Figure shows that X-ray isophotes are more flattened in cluster
cores in the CSF runs compared to the NR clusters. There is a clear
rapid upturn in $\epsilon_{\rm CSF}$ at $r\lesssim 0.1r_{500}$
reflecting rotational motions of gas in these runs. Note that we do
not confirm results of F09 who claimed significant
flattening of isophotes due to rotation out to $r\approx 0.4r_{500}$
based on the analysis of the same simulations we use in this study. 

There is a downturn in $\epsilon$ for the 
NR clusters at similar radii ($r \lesssim 0.1r_{500}$) 
reflecting the effects of random gas motions. These trends are
consistent with the results for the three-dimensional distributions
presented in Section~\ref{sec:3d_results} and are just as
pronounced. The difference between CSF and NR runs is particularly
large for relaxed clusters, which have had more time since their most recent 
major mergers during which gas cooling could proceed.

Outside the cluster cores, $0.1<r/r_{500}<1$, the ellipticities are 
approximately constant in both the CSF and NR runs.  The average ellipticities 
in this range differ for the two sets of relaxed clusters.  
Relaxed clusters in the CSF simulations yield an average ellipticity of 
$\langle\epsilon_{\rm CSF}\rangle \approx 0.1-0.15$ while the relaxed NR 
clusters have $\langle\epsilon_{\rm NR}\rangle \approx 0.15$-$0.25$.  
The more spherical DM distribution and shape of the isopotential contours in the CSF
clusters compared to the NR clusters are clearly discernible in X-ray maps of 
the relaxed systems.  Note that these differences are highly significant both in 
individual bins for relaxed clusters, and because they persist over a wide range of 
radii.  The average ellipticities of
unrelaxed clusters in the CSF and NR clusters are 
$\langle\epsilon_{\rm CSF}\rangle \approx 0.25$ and
$\langle\epsilon_{\rm NR}\rangle \approx 0.25-0.35$ respectively. 
The mean cluster ellipticities and their $1\sigma$ errors on the mean are summarized in Table~\ref{tab:ellipticity}. 
Note that the scatter is substantial and constraining the net effect of cooling and the prevalence 
of random motions will require averaging over a sample of relaxed clusters.

\section{Summary and Discussion}
\label{sec:discussions}

We have investigated the axis ratios of DM mass, 
hot gas, and gravitational potential using 16 clusters simulated from the same initial conditions, 
but with different baryonic physics (with and without radiative cooling and star formation).  
Our results can be summarized as follows.

\begin{itemize}
\item We show that gas distribution in simulated clusters has a rather spherical shape at large cluster-centric radii with the average axis ratios of $\sim 0.8$ at $r\gtrsim 0.5r_{500}$. This implies that the standard assumption of spherical symmetry in analyses of ICM using X-ray and Sunyaev-Zeldovich observations should be quite accurate.  
\item We show that baryonic dissipation makes gas distribution more spherical 
at $0.1\lesssim r/r_{500}\lesssim 1$, where the axis ratios are larger by 
$\sim 0.2$ on average in the cooling (CSF) runs compared to the non-radiative (NR) 
runs.  At small radii $r/r_{500} \lesssim 0.1$, 
the short-to-long axis ratios $c/a$ are lower in the CSF runs compared to the NR runs 
by as much as $\sim -0.5$, but the intermediate axis ratios $b/a$ are reduced to a 
much smaller degree.  The CSF gas distributions are oblate in their centers 
reflecting the presence of cool gas supported by rotation.  

\item We present predictions for X-ray ellipticity profiles of 
intracluster gas based on mock {\em Chandra} X-ray maps of our simulated clusters.  
We show that the effect of cooling on the ellipticity of gas is consistent with the 
three-dimensional results.  Specifically, the ellipticities at larger radii 
($0.1\lesssim r/r_{500c}\lesssim 1$) decrease for the CSF clusters compared to NR 
clusters reflecting the rounder gravitational potential of the CSF clusters.  
At $r \lesssim 0.1r_{500}$, ellipticities in the CSF clusters increase with decreasing radius.   

\item NR clusters exhibit the opposite trends.  NR clusters are more triaxial than the CSF clusters 
at $r\gtrsim 0.1r_{500}$, but become considerably rounder at $r \lesssim 0.1r_{500}$. The latter trend 
is due to random gas motions, which are present at all radii but become considerable compared to 
the thermal energy of the gas (as reflected by increasing $\sigma_{\rm gas}/v_{\rm circ}$ ratio).  

\item Our results indicate that observed ellipticity profiles of X-ray clusters can be used to 
constrain both the amount of cooling in the last several billion years of cluster evolution and 
presence of significant random gas motions. 
\end{itemize}

There are several issues to bear in mind when interpreting our
results. First is the issue of overcooling in galaxy formation 
simulations.  Our CSF simulations
suffer from overcooling in the cluster cores ($r \lesssim 0.1r_{500}$),
such that the effect of halo contraction in response to the formation
of the central galaxy is likely overestimated. The dissipational
effect on gas shapes that we present can therefore be considered as an
upper limit. At the same time, some amount of cooling must have
occurred because galaxies do exist in clusters and the
observed ICM gas fractions are significantly below the expected values
\citep[e.g.,][]{kravtsov_etal09}. We therefore expect the
ellipticity of the intracluster gas in real clusters to be between our
results for CSF and NR runs.

Although our cluster sample is not drawn to sample the mass function of clusters, 
the mass-dependence of ellipticities is very weak (Figure~\ref{fig:sm_z0}), 
at least in the range of masses we probe 
($3.5\times10^{13}\,h^{-1}M_{\odot}<M_{500}<9\times10^{14}\,h^{-1}M_{\odot}$) 
and our results should therefore not be biased significantly. 

A more significant concern is the possible bias due to the fact that
our cluster sample was simulated assuming $\sigma_8 =0.9$, while the
most recent estimates indicate $\sigma_8= 0.80\pm0.02$
\citep{vikhlinin_etal09,jarosik_etal10}.  The effect of a lower
$\sigma_8$ is that halos would form later, leading to higher
ellipticities \citep{allgood_etal06}.  However, \citet{maccio_etal08} show that 
effect of changing $\sigma_8$ from $0.9$ to $0.8$ changes average $c/a$ ratios of 
halos by only $\approx 0.03$, considerably smaller than the differences discussed 
in this paper.  Moreover, both the CSF and NR clusters would be affected by the 
difference in cosmology and it is plausible that they would be affected to a 
similar degree. 

Another potential concern is that AGN heating is not included in our
simulations though evidence of such heating is abundant in real clusters, 
especially in relaxed cool-core clusters \citep[e.g.,][]{mcnamara_nulsen07}.  Jets and
bubbles inflated by the AGNs can potentially change the shape of the
gas significantly in the cluster core.  We expect that the AGN would
also change the gas shape in the cluster outskirts by suppressing the 
overall amount of cooling throughout the cluster formation.

Despite these caveats, a few general implications can be drawn from
our results. First, as dissipation makes the gas more spherical,
systematics of observable quantities integrated along the line of
sight such as $Y_{\rm SZ}$, or any deprojected quantity that relies on the
assumption of spherical symmetry, should be reduced compared to
conclusions one could draw from dissipationless simulations.  Another
implication is that the shape of gas can be different from the shape
of the potential in the cluster core $r<0.1r_{500}$, and in the
cluster outskirts $r>r_{500}$. The difference in shape between gas and
potential can be attributed to deviation from hydrostatic equilibrium
as shown in Section~\ref{subsec:gas_vs_pot}. In the intermediate
radial range $0.1\leq r/r_{500}\leq 1$ , however, we have shown that
the gas shape generally coincides with that of the potential, and
therefore the shape of the potential can be inferred from the shape of
gas.  If there is an independent way of determining the shape of
gravitational potential, e.g. by gravitational lensing, one may be
able to constrain the amount of gas motions by comparing the shape of
the gravitational potential and the shape of gas.

However, even without independent information about the potential, our results indicate 
that ellipticities derived from X-ray images alone can constrain the amount of cooling and 
the presence of random gas motions. 
This is because these effects result in a rapid change of ellipticity with decreasing radius at $r<0.1r_{500}$ but of 
an opposite sign.   We will present comparisons of gas ellipticities in simulations and observations 
from {\em Chandra} and {\em ROSAT} in a separate companion paper ({E.} {T.} Lau et al. 2011, in preparation).   

\acknowledgments 
We thank the referee David Buote for constructive comments on this paper, and for sending 
us the FITS file they used in their paper. We would also like to thank Alexey Vikhlinin for helpful comments. 
EL and AK are supported 
by the NSF grant AST-0708154, by NASA grant NAG5-13274, 
and by Kavli Institute for Cosmological Physics at the University of Chicago 
through grant NSF PHY-0551142 and an endowment from the Kavli Foundation.  
DN acknowledges the support of Yale University.  The work of ARZ is funded by the 
University of Pittsburgh, the NSF through grant AST-0806367 and by the DOE. 
The cosmological simulations used in this
study were performed on the IBM RS/6000 SP4 system (copper) at the
National Center for Supercomputing Applications (NCSA). This work made
extensive use of the NASA Astrophysics Data System and arXiv.org
preprint server.

\bibliographystyle{apj}
\bibliography{ms}

\end{document}